\documentclass[pra,twocolumn,showpacs,preprintnumbers,superscriptaddress]{revtex4}
\usepackage{times}
\usepackage{bm}
\usepackage{graphicx}
\usepackage{amsbsy}
\usepackage{amsmath}
\usepackage{amsfonts}
\usepackage{amsthm}
\usepackage{xcolor}

\begin{document}
	\theoremstyle{plain}
	\newtheorem{theorem}{Theorem}
	\newtheorem{lemma}[theorem]{Lemma}
	\newtheorem{corollary}[theorem]{Corollary}
	\newtheorem{proposition}[theorem]{Proposition}\newtheorem{conjecture}[theorem]{Conjecture}
	\theoremstyle{definition}
	\newtheorem{definition}[theorem]{Definition}
	\theoremstyle{remark}
	\newtheorem*{remark}{Remark}
	\newtheorem{example}{Example}
	\title{Entanglement detection in arbitrary dimensional bipartite quantum systems through partial  realigned moments}
	\author{Shruti Aggarwal$^1$, Satyabrata Adhikari$^1$, A. S. Majumdar$^2$}
	\email{shruti_phd2k19@dtu.ac.in, satyabrata@dtu.ac.in, archan@bose.res.in} \affiliation{$^1$Delhi Technological University, Shahbad Daulatpur, Main Bawana Road, Delhi-110042,
		India,\\ $^2$S. N. Bose National Centre for Basic Sciences, Block JD, Sector III, Salt Lake, Kolkata 700098, India}
\begin{abstract}
Detection of entanglement through partial knowledge of the quantum state is a challenge to implement efficiently. Here we propose a separability criterion for detecting bipartite entanglement in arbitrary dimensional quantum states using partial moments of the realigned density matrix. Our approach
enables the detection of both distillable and bound entangled states through a common framework. We illustrate the significance of our method through examples of states belonging to both the above categories, which are not detectable using comparable other schemes relying on partial state information. The formalism of employing partial realigned moments proposed here is further
shown to be effective for two-qubit systems too, with a slight modification of our separability criterion.
\end{abstract}
	 \maketitle
	\section{Introduction}
	Entanglement \cite{epr, schrod} is a remarkable feature of quantum systems with no classical analogue. It is the key physical resource to realize quantum information	tasks such as quantum teleportation \cite{tele}, quantum cryptography \cite{cryp}, and quantum dense coding \cite{niel}. This has provided a strong motivation for the characterization and classification of entangled states. In particular, criteria to decide whether or not a given quantum state is entangled are of high theoretical and practical interest. Even though numerous entanglement criteria have been proposed in the past years, nevertheless there exists no universally applicable method to tell us whether a given quantum state is entangled or not. Historically,
Bell-type inequalities were the first operational criterion
to distinguish between entangled and separable states \cite{bell}. Due to the importance of entanglement in quantum information processing, there has been a steady quest for devising more and more efficient methods for detecting entanglement in quantum states \cite{Oguhne,rhoro}. 

For bipartite systems, there exist two famous separability
criteria: the positive partial transpose (PPT) criterion \cite{peres} and the matrix realignment criterion \cite{chenwu, orud}. The former criterion can detect the entanglement of all nonpositive partial transpose (NPT) states but cannot detect any PPT entangled state. The latter criterion is weaker than the former one over NPT states; however, it can detect some PPT entangled states. Apart from the above, there exist operational tools to detect entanglement in practice. Entanglement witness based on measurement of observables provides a method to characterize entanglement \cite{terhal,guhne2003}. 
There exist different schemes for the construction of witness operators in the literature \cite{lewen,ganguly,shruti1,shruti2}, though all such schemes rely on certain prior information about the quantum state. Besides, entanglement can also be detected using measurement statistics in a device independent manner through an approach called self-testing \cite{selftest, selftest2, selftest3, selftest4} which again relies on certain additional assumptions. \\

In practical situations, complete information about the quantum state may not be always available, and entanglement detection based on the partial knowledge of the density matrix may be
easier to implement in experiments \cite{entdet}. Recently, Elben et al.  \cite{elben} proposed a method for detecting bipartite entanglement based on estimating moments of the partially transposed density matrix. Nevel et al.  \cite{neven} proposed an ordered set of experimentally accessible conditions for detecting entanglement in mixed states. Moments have the advantage that they can be estimated using shadow tomography in a more efficient way than if one had to reconstruct the state via full quantum state tomography.
Such works \cite{elben,neven,guhne,lin} are focussed on the detection of NPT entangled states only.   Detection of bound entangled states using moments needs a deeper investigation. In the present paper, we provide a separability criterion based on moments of the Hermitian matrix obtained after applying the realignment operation for quantum states in arbitrary dimensional bipartite systems.

The motivation of our work is to construct  computable entanglement conditions that can detect NPT entangled states as well as bound entangled states (BES) through partial knowledge of the density matrix. Recently, detection of bound entangled states using a moment based criterion has been studied in \cite{tzhang, gamma}.  Our criterion based on moments is stronger in the sense that lower-order realigned moments are sufficient to detect entanglement. Importantly, this approach is not only conceptually sound but also tractable from an experimental perspective. The classical shadows formalism allows for reliably estimating moments from randomized single-qubit measurements \cite{chad,kett}. If we make multiple copies of a state represented by an $m \times m$ density matrix $\rho$, the moments $tr(\rho)^2$,\;.\;.\;.\;, $tr(\rho)^{m}$ can be measured using cyclic shift operators  \cite{horoekert,keyl}. It has been shown that measuring partial moments is technically possible using $m$ copies of the state and controlled swap operations \cite{ha, cai, barti}. A method based on machine learning for measuring moments of any order has also been proposed \cite{saugato}.

In this paper, we introduce a novel entanglement criterion for bipartite systems based on the moments of the hermitian matrix  $({\rho^R})^{\dagger} \rho^R$ obtained after applying realignment operation on a quantum state $\rho$, which we call here realigned moments (or $R$-moments). First, we show that our criterion effectively detects bound entangled states (BES) in higher dimensional two particle systems. Further, we show that it performs better than the other existing criteria based on partial moments in some cases for the detection of NPT entangled states in higher dimensional systems. The $R$-moment criterion is formulated as a simple inequality that must be fulfilled by separable states of bipartite systems, and hence, its violation by a state reveals its entanglement.  We illustrate the significance of the $R$-moment criterion for the detection of NPT and bound entanglement by examining some examples. Finally, we devise another separability criterion for 2-qubit systems and probe some examples of 2-qubit NPT entangled states that are undetected by other criteria using partial moments and realignment. 

The paper is organized as follows. In Sec. II, we briefly review some entanglement detection criteria. In Sec. III, we present the idea of using realigned moments ($R$-moments) for entanglement detection. In Sec. IV. We discuss the $R$-moment criterion for the detection of higher dimensional bipartite entangled states and demonstrate with examples that it can detect both NPT as well as bound entangled states effectively. In Sec. V, we propose another criterion based on $R$-moment for $2 \otimes 2$ dimensional systems. We illustrate the utility of the $R$-moment criterion by providing some examples and also by comparing it with other entanglement detection criteria based on partial moments. In Sec. VI we propose a scheme for the measurement of moments of the realigned matrix. We conclude in Sec. VII with a brief discussion.
	
\section{Preliminaries}
Let $\mathcal{H}= {\mathcal{H}_{A}^{m}} \otimes {\mathcal{H}_{B}^{n}}$ denote the composite Hilbert space of dimension $mn$, where the subsystem $A$ has dimension $m$ and the subsystem $B$ has dimension $n$. A quantum state $\rho \in \mathcal{D(H)}$ is said to be separable if it can be expressed as a convex combination of product states of two subsystems,
\begin{equation}
	\rho = \sum_i p_i \rho^A_i \otimes \rho^B_i ,\; \sum p_i =1,\; 0 \leq p_i \leq 1
	\label{sepform}
\end{equation}
Otherwise, the state is called entangled.
The detection of an entangled state using the above definition is not feasible in practice because one has to verify whether the state under investigation can be written in the form (\ref{sepform}) for every basis existing in the composite Hilbert space. Thus, a lot of research is devoted to developing different criteria that may detect the entangled states. We discuss here a few of them which may be considered to be efficient criteria. 

\subsection{PPT criterion}
 The first entanglement detection criterion was established by Peres \cite{peres} and later it was proved to be necessary and sufficient for $2 \otimes 2$ and $2 \otimes 3$ systems \cite{horo}. This criterion is popularly known as the PPT criterion. It states that for any bipartite separable state $\rho$, partial transposition with respect to subsystem $B$ denoted by  $\rho^{\tau_B}$ with elements $\rho^{\tau_B}_{ij,kl} = \rho_{il,kj}$ is also a density operator; i.e., it has non-negative eigenvalues. The states that satisfy the PPT criteria are known as PPT states. In higher dimensional systems, it has been observed that there are PPT states which are entangled. Therefore, the PPT criterion is only a necessary condition but not sufficient for separability for higher dimensional systems. PPT entangled states are also called bound entangled states since they cannot be distilled into any maximally entangled pure singlets \cite{phoro}. Partial transposition is positive but not completely positive and hence it may not be realized physically in an experiment \cite{horoekert}. Horodecki et al \cite{horoekert} proposed a method to modify the partial transposition operation so that it becomes experimentally viable without involving any state estimation. This method is called structural physical approximation (SPA). SPA offers a systematic way of approximating those non-physical maps that are positive but not completely positive, with quantum channels.

\subsection{Realignment criterion}
Like partial transposition, one may use another permutation of the density matrix elements. One such permutation of the density matrix elements is used by K. Chen et al \cite{kchen} and O. Rudolph \cite{rudolph} to study the entanglement detection problem. The map used to obtain the permutation is known as the realignment map. This map may be used to obtain another separability criterion for bipartite systems, which is known as the realignment criterion. It states that for any bipartite separable state $\rho$, the realigned matrix $\rho^R$ with elements $\rho_{ij,kl} = \rho_{ik,jl}$ has trace norm not greater than 1, i.e., $||\rho^R||_1 \leq 1$, where the trace norm of an operator $X$ is defined as $||X||_1 := Tr(\sqrt{X^{\dagger}X)}$ \cite{chenwu,li}. This criterion provides us only a necessary condition for lower as well as higher dimensional systems. Nevertheless, this criterion is considered to be a strong criterion since it can detect NPT entangled states as well as bound entangled states. Furthermore,    an analytical lower bound of the concurrence of arbitrary dimensional bipartite states based on the realignment criterion
was also derived \cite{kchen}. It makes it possible to estimate the amount of entanglement in bound entangled states too. Separability criteria based on the realignment
of density matrices and reduced density matrices have
been proposed in \cite{qshen}. In \cite{xqi}, the rank of the realigned matrix is used to obtain necessary and sufficient product criteria for quantum states. Recently, it has been shown that realignment criteria may be realized physically using structural physical approximation \cite{shruti3}.

\subsection{Entanglement detection criteria using  moments}
 Elben et al. \cite{elben} proposed a method for detecting bipartite entanglement in a many-body mixed state based on estimating moments of the partially transposed density matrix. Since partial transposition operation is a positive but not completely positive map, it is not physical and thus it may not be implemented in the experiment. But despite the above difficulty in realizing the partial transposition operation in the experiment, the measurement of their moments is possible \cite{elben}.  A condition to detect entanglement called $p_3$-PPT criterion, was proposed using the first three PT moments. If $\tau$ denotes the partial transposition operation, then the $k^{th}$ partial moment is defined as 
 \begin{eqnarray}
  p_k(\rho^{\tau})=Tr[(\rho^{\tau})^k] 
\end{eqnarray}  
 PT moment has the advantage that it can be estimated using shadow tomography in a more efficient way than if one had to reconstruct the state $\rho$ via full quantum state tomography \cite{elben}.
The $p_3$-PPT condition states that any PPT state $\rho$ satisfies the following inequality
\begin{equation}
	{L}_1 \equiv (p_2(\rho^{\tau}))^2 - p_3(\rho^{\tau}) p_1(\rho^{\tau}) \leq 0 \label{p3ppt}
\end{equation}
The violation of the above inequality (\ref{p3ppt}) by any $d_1 \otimes d_2$ dimensional bipartite state $\rho$ indicates that the state is an NPT entangled state.

Neven et al. \cite{neven} proposed a set of inequalities, known as $D_k^{(in)}$ inequalities to detect bipartite NPT entangled states. Each $D_k^{(in)}$ involves the first $k$ moments of the partially transposed operator $\rho^{\tau}$. The violation of any single $D_k^{(in)}$ inequality implies $\rho$ is NPT entangled state.
One can obtain the first nontrivial condition in the form of $D_3^{(in)}$ inequality, which reads
\begin{equation}
	{L}_2 \equiv \frac{3}{2} (p_1(\rho^{\tau}))(p_2(\rho^{\tau})) - \frac{1}{2}(p_1(\rho^{\tau}))^3 - p_3(\rho^{\tau})  \leq 0
	\label{d3}
\end{equation}
The inequality (\ref{d3}) is satisfied by all PPT states and its violation certifies the existence of an NPT entangled state, which may be expressed as 
\begin{equation}
	{L}_2  > 0.
	\label{d3v}
\end{equation}
Neven et al. \cite{neven} showed that knowing only the first three moments $p_1(\rho^{\tau})$, $p_2(\rho^{\tau})$ and $p_3(\rho^{\tau})$, the above inequality   detects more entangled states than the $p_3$-PPT criterion when the purity of $\rho^{\tau}$ is greater than or equal to $1/2$, i.e., when $1/2 \leq p_2(\rho^{\tau}) \leq 1$. In the other region, i.e., when $0 \leq p_2(\rho^{\tau}) < 1/2$, the $p_3$-PPT criterion detects more entangled states than the $D_3^{(in)}$ criterion.\\
Yu et al. \cite{guhne} introduced an optimal entanglement detection criteria based on partial moments called $p_3$-OPPT criterion. This optimal separability criterion can be stated as follows. If $\rho$ is separable, then the following inequality holds:
\begin{eqnarray}
	{L}_3 =	\mu x^3 + (1-\mu x)^3 - p_3(\rho^{\tau}) \leq 0
	\label{l3}
\end{eqnarray}
where $x= \frac{\mu + \sqrt{\mu [p_2(\rho^{\tau})(\mu + 1 )-1]}}{\mu(\mu+1)}$ and $\mu = \lfloor \frac{1}{p_2(\rho^{\tau})} \rfloor$. \\


Zhang et al. \cite{tzhang} proposed another entanglement detection
criteria in terms of the quantities called realignment moments. To derive their entanglement detection criterion, they have defined the realignment moments for a $d \otimes d$ dimensional bipartite state $\rho$ as 
\begin{eqnarray}
	r_k (\rho^R) = Tr[\left(\rho^R (\rho^R)^{\dagger}\right)^{k/2}],\; k = 1, 2, . . ., d^2
\end{eqnarray}
where $d^2$ is order of the matrix $\rho^R$.\\
The separability criterion based on realignment moments $r_2$ and $r_3$ is stated as follows. If a quantum state $\rho$ is separable, then
\begin{eqnarray}
	{L}_4 \equiv (r_2(\rho^R))^2 - r_3(\rho^R) \leq 0	\label{rzhang}
\end{eqnarray}
Violation of the inequality (\ref{rzhang}), i.e., ${L}_4 > 0$ implies that the state $\rho$ is entangled.

A stronger separability criterion based on Hankel matrices and involving higher order $r_k$ has been derived in \cite{tzhang}.
For $r=(r_0, r_1, r_2, . . . r_n)$, Hankel matrices can be constructed as  $[{H}_k (r)]_{ij} = r_{i+j}$ and $[{B}_l (r)]_{ij}=r_{i+j+1}$ for $i,j=0,1,2,. . ., k$. The criterion may be stated as follows.
If $\rho$ is separable, then for $k = 1, 2, . . ., \lfloor \frac{n}{2} \rfloor$  and $l = 1, 2, . . ., \lfloor \frac{n - 1}{2} \rfloor$, we have 
\begin{eqnarray}
	\widehat{H}_k (r) &=& [r_{i+j}(\rho^R)] \geq 0 \label{hk} \\ \widehat{B}_l (r)&=& [r_{i+j+1} (\rho^R)] \geq 0
	\label{bl}
\end{eqnarray}
with $r_1(\rho^R) = 1$.


\section{Realigned moments or $R-$moments}
 Before presenting our separability criterion based on realigned moments, let us first define the idea of realigned moments or $R$-moments in $m \otimes n$ dimensional systems.  To make the task simpler, consider first a $2\otimes 3$ system described by the density operator $\sigma_{12}$ which is given by
 \begin{eqnarray}
 	\sigma_{12}= 
 	\begin{pmatrix}
 		Z_{11} & Z_{12} \\
 		Z_{21} & Z_{22} 
 	\end{pmatrix}
 \end{eqnarray}
 where $Z_{11}=\begin{pmatrix}
 	t_{11} & t_{12} & t_{13} \\
 	t_{12}^{*} & t_{22} & t_{23} \\
 	t_{13}^{*} & t_{23}^{*} & t_{33} \\
 \end{pmatrix}$, $Z_{12}=\begin{pmatrix}
 	t_{14} & t_{15} & t_{16} \\
 	t_{24} & t_{25} & t_{26} \\
 	t_{34} & t_{35} & t_{36} \\
 \end{pmatrix}$, $Z_{21}=Z_{12}^{\dagger}$, $Z_{22}=\begin{pmatrix}
 	t_{44} & t_{45} & t_{46} \\
 	t_{45}^{*} & t_{55} & t_{56} \\
 	t_{46}^{*} & t_{56}^{*} & t_{66} \\
 \end{pmatrix}$.\\
 The normalization condition of $\sigma_{12}$ is given by $\sum_{i=1}^{6}t_{ii}=1$.
 The realigned matrix of $\sigma_{12}$ is denoted by $\sigma_{12}^{R}$ and it is given by
 \begin{eqnarray}
 	\sigma_{12}^{R}&=& \begin{pmatrix}
 		(vecZ_{11})^{T} \\
 		(vecZ_{12})^{T}\\
 		(vecZ_{21})^{T}\\
 		(vecZ_{22})^{T} 
 	\end{pmatrix}\nonumber\\&=&\begin{pmatrix}
 		t_{11} & t_{12} & t_{13} & t_{12}^{*} & t_{22} & t_{23} & t_{13}^{*} & t_{23}^{*} & t_{33}\\
 		t_{14} & t_{15} & t_{16} & t_{24} & t_{25} & t_{26} & t_{34} & t_{35} & t_{36}\\
 		t_{14}^{*} & t_{24}^{*} & t_{34}^{*} & t_{15}^{*} & t_{25}^{*} & t_{35}^{*} & t_{16}^{*} & t_{26}^{*} & t_{36}^{*} \\
 		t_{44} & t_{45} & t_{46} & t_{45}^{*} & t_{55} & t_{56} & t_{46}^{*} & t_{56}^{*} & t_{66} \\
 	\end{pmatrix}
 \end{eqnarray}
 where for any $n \times n$ matrix $X_{ij}$ with entries ${x_{ij}}$, $vecX_{ij}$ is defined as
 \begin{eqnarray}
 	vecX_{ij} = [x_{11}, . . ., x_{1n}, x_{21}, . . ., x_{2n},. . .,x_{n1}, . . . x_{nn}]^T
 \end{eqnarray}
 
 Note that $(\sigma_{12}^{R})^{\dagger}\sigma_{12}^{R}$ is a matrix of order $9\times 9$. 
 Also, the number of non-zero singular values of $\sigma_{12}^R$ is equal to the rank of $\sigma_{12}^R$ and hence it will be at most four. 
 We can now generalize this fact for $m\otimes n$ systems. 
 Let $\rho$ be a density matrix representing a $m \otimes n$ dimensional state and it can be written as a block matrix with $m$ number of blocks in each row and column with each block being a $n \times n$ matrix.
 The realigned matrix $\rho^R$ obtained after applying the realignment operation has dimension $m^2 \times n^2$. 
 The first step to obtain the $R$-moment criterion is to find the characteristic equation of the $n^2 \times n^2$ Hermitian operator $({\rho^R})^{\dagger} \rho^R$. It is given by
 \begin{eqnarray}
 	&&det({\rho^R}^{\dagger} \rho^R-\lambda I) = 0\nonumber\\&&
 	\implies \prod_{i=1}^{n^2} (\lambda_i({\rho^R}^{\dagger} \rho^R) - \lambda)=0\nonumber\\&&
 	\implies  \lambda^{n^2} + D_1 \lambda^{n^2-1} + D_2\lambda^{n^2-2} + . \; .\;. \;. + D_{n^2} = 0  \nonumber \\ \label{charr}
 \end{eqnarray}
 where $\lambda_i({(\rho^R)}^{\dagger} \rho^R)$ given in the second step denotes the roots of the characteristic polynomial (\ref{charr}). 
 Using well-known results related to Newton polynomials and the Faddeev-LeVerrier algorithm for the characteristic polynomial and traces of powers of a matrix \cite{verrier,faddeev,zeil},
 the coefficients $\{D_i\}_{i=1}^{n^2}$ given in the third step can be described in terms of moments of ${(\rho^R)}^{\dagger} \rho^R$, i.e., 
 \begin{eqnarray}
 	D_i = (-1)^i \frac{1}{i!} 
 	\begin{vmatrix}
 		T_1 & T_2 & T_3 &.&.&.&...&T_m \\
 		1 & T_1 & T_2 & T_3 &.&.&...&T_{m-1} \\
 		0 & 2 & T_1 & T_2 & T_3 &.&...&T_{m-2} \\
 		0 & 0 & 3 & T_1 & T_2 & T_3 &...&T_{m-3} \\
 		. & . & . & . & . & . &...&. \\
 		. & . & . & . & . & . &...&. \\
 		. & . & . & . & . & . &\ddots&. \\
 		0 & 0 & 0 & 0 & 0 & 0 &...& T_1 \\
 	\end{vmatrix} \label{dm}
 \end{eqnarray}
 for $i=1$ to $n^2$, where $D_{n^2} = det({(\rho^R)}^{\dagger} \rho^R)$. $T_k = tr(({(\rho^R)}^{\dagger} \rho^R)^k)$ denotes the $k^{th}$ realigned moment of ${(\rho^R)}^{\dagger} \rho^R$.
 
 Let us arrange the singular values of $\rho^R$ in descending order, i.e., $\sigma_1(\rho^R) \geq . \; . \; . \geq \sigma_r(\rho^R) \geq 0$, where $r$ denotes the total number of singular values of $\rho^R$. 
 Thus for $i=1$ to $r$, we have 
 $$ \lambda_i({\rho^R}^{\dagger} \rho^R) = \sigma_i^2(\rho^R)$$  Therefore, the relation between the coefficients $D_{i}$ and the singular values  $\sigma_i(\rho^R)$ is given by
 \begin{eqnarray}
 	\sum_{i=1}^r \sigma_{i}^2 (\rho^R) &=& -D_1 \label{re1} \\
 	\sum_{i < j} \sigma_{i}^2 (\rho^R) \sigma_{j}^2 (\rho^R) &=& D_2 \label{e2}\\
 	\sum_{i < j < k} \sigma_{i}^2 (\rho^R) \sigma_{j}^2 (\rho^R) \sigma_{k}^2 (\rho^R)&=& -D_3 \label{D3} \\
 	....&&.... \nonumber\\
 	....&&....\nonumber\\
 	....&&.... \nonumber\\
 	\prod_{i=1}^{r}	\sigma_{i}^2 (\rho^R)  &=& D_r \label{e4}
 \end{eqnarray}
 Using (\ref{dm}), the coefficients $D_1, D_2$ and $D_3$ in terms of first moment $T_{1}$, second moment $T_{2}$ and third moment $T_{3}$ of ${\rho^R}^{\dagger}\rho^R$ can be expressed as
 \begin{eqnarray}
 	&&D_1 = -T_1 \label{d1}\label{rd_1} \\&&
 	D_2 = \frac{1}{2} ({T_1}^2 - T_2)\label{d_2}\\&&
 	D_3 = -\frac{1}{6} ({T_1}^3 - 3T_1 T_2 + 2 T_3) \label{d_3}
 \end{eqnarray}
 In general, the coefficient $D_{n^2}$ may be expressed as $D_{n^2} = det({\rho^R}^{\dagger} \rho^R)$.



\section{Separability criterion based on $R$-moments in $m \otimes n$ systems}
 Although the separability criteria based on partial moments involve up to $3^{rd}$ order moments \cite{elben,neven}, it fails to detect several NPT entangled states in higher dimensional systems and is also not applicable for the detection of bound entangled states.   This gives us a strong motivation to investigate the concept of realigned moments in entanglement detection in higher dimensional systems. 
  
  \subsection{Criterion}
 We are now ready to present our separability criterion based on realigned moments for  $m \otimes n$ dimensional systems. It is formulated in the form of an inequality that involves the $k^{th}$ order moments where $k$ is the rank of the matrix $\rho^R$.

\textbf{Theorem 1:} Let $\rho$ be any bipartite state in $m \otimes n$ dimensions. Consider the $k$ non-zero singular values of the realigned matrix $\rho^R$ that may be denoted as $\sigma_1, \sigma_2, \ldots \sigma_k$ with $1\leq k \leq min\{m^2,n^2\}$. If $\rho$ is separable then the following inequality holds:
\begin{equation}
	{R}_1 \equiv	k(k-1) {D_k}^{1/k} + T_1 - 1 \leq 0 \label{thm4}
\end{equation}
where $D_k = \prod_{i=1}^{k} \sigma_i^2(\rho^R)$ and $T_1 = Tr[(\rho^R)^{\dagger} \rho^R]$.\\
\textit{Proof:} Let $\rho$ be any separable state in $m \otimes n$ system and assume that the number of non-zero eigenvalues of $(\rho^{R})^{\dagger}\rho^{R}$ are $k$.  That is, there exist a number $k$  $(1\leq k \leq min\{m^2,n^2\})$ depending upon the number of non-zero singular values of $\rho^R$ for which $D_{k}=\prod_{i=1}^{k} \sigma_i^2(\rho^R)\neq 0$.
The degenerated characteristic equation of $(\rho^R)^{\dagger} \rho^R$ is given by
\begin{eqnarray}
	\lambda^k + \sum_{i=1}^{k} D_i \lambda^{k-i} = 0
\end{eqnarray}
where $D_i$ is defined in (\ref{dm}) for $i=1$ to $k$ and $D_i = 0$ for $i>k$.

Using (\ref{re1}) and (\ref{rd_1}), the first realigned moment $T_1$ can be expressed  in terms of the singular values of $\rho^R$ as
\begin{eqnarray}
	T_1 &=&\sum_{i=1}^{k} \sigma_{i}^2 (\rho^R) \nonumber\\ &=& \left(\sum_{i=1}^{k} \sigma_{i} (\rho^R)\right)^2 - 2\sum_{i < j} \sigma_{i} (\rho^R) \sigma_{j} (\rho^R)  \label{e5}
\end{eqnarray}
(\ref{e5}) can be re-expressed as
\begin{eqnarray}
	\sum_{i < j} \sigma_{i} (\rho^R) \sigma_{j} (\rho^R)= \frac{1}{2} \left(\left(\sum_{i=1}^{k} \sigma_{i} (\rho^R)\right)^2 - T_1\right) \label{e6}
\end{eqnarray}
It is elementary to note that the arithmetic mean of a list of non-negative real numbers is greater than or equal to their geometric mean.
Since $\sigma_i$'s for $i=1$ to $k$ are non-negative real numbers, so we have 
\begin{eqnarray}
	\sum_{i < j} \sigma_{i} (\rho^R) \sigma_{j} (\rho^R) \geq \frac{k(k-1)}{2}\left(\prod_{i=1}^{k} {\sigma_i}(\rho^R)\right)^{2/k} \label{e8}
\end{eqnarray}

Using (\ref{e6}) and (\ref{e8}), we get
\begin{eqnarray}
	\frac{1}{2} (||\rho^R||_1^2 - T_1) \geq \frac{k(k-1)}{2} D_k^{1/k} \label{e9}
\end{eqnarray}
where $||\rho^R||_1 =\sum_{i=1}^{k} \sigma_{i} (\rho^R)$ and  $D_k= \prod_{i=1}^{k} \sigma_i^2(\rho^R)$.
Thus, we obtain 
\begin{equation}
	||\rho^R||_1 \geq k(k-1) D_k^{1/k} +T_1 \label{e10}
\end{equation} 
Since $\rho$ is any arbitrary separable state, using the realignment criterion in the above inequality (\ref{e10}), we have $k(k-1) D_k^{1/k} +T_1 \leq 1$, which proves (\ref{thm4}).

\textbf{Corollary 1:} \label{cor3} Let $\rho$ be any bipartite state in $m \otimes n$  dimensional system. Let $\sigma_1, \sigma_2, \ldots \sigma_k$ with $1\leq k \leq min\{m^2,n^2\}$ be $k$ non-zero singular values of the realigned matrix $\rho^R$. If any state $\rho$ violates (\ref{thm4}), then it is an entangled state.

It is to be noted that the $R$-moment based separability criterion we have developed here is more fruitful for those density matrices $\rho$ in $m \otimes n$  system for which $det((\rho^R)^{\dagger} \rho^R) = 0$.  Therefore, the $R$-moment criterion works well when $\rho^R$ is non-full rank. To test the separability criteria based on $R$-moments, consideration of the non-full rank state is advantageous in the sense that it does not require all the $R$-moments, and hence, our criterion holds good even when we do not have full information of the state.   In $m \otimes n$ system, the condition $det((\rho^R)^{\dagger} \rho^R) = 0$ is valid when  (i) the number of non-zero eigenvalues of the matrix $(\rho^{R})^{\dagger}\rho^{R}$  are less than $n^{2}$ where $m \geq n$, or when (ii) the number of non-zero eigenvalues of the matrix $(\rho^{R})^{\dagger}\rho^{R}$ are less than $m^{2}$ where $m \leq n$. In particular, 
for any $2 \otimes 2$ system, the separability criterion given in Theorem 1 reduces to 
\begin{equation}
	12 {D_4}^{1/4} + T_1 \leq 1 \label{thm5}
\end{equation}
The violation of the inequality (\ref{thm5}) by any 2-qubit state indicates the fact that the state is entangled. 

It may be noted
that there
is no universal entanglement detection criterion that could outperform all
other criteria. Any chosen criterion could work better for a given class
of states, and vice-versa. 
We now discuss  examples of PPT and NPT entangled states in $3\otimes 3$ and $4 \otimes 4$ systems which are detected by our $R$-moment criterion, but 
not by certain other criteria contained in the literature. 

\subsection{Examples}
In this section, we have considered a few examples of bipartite two-qutrit, two-ququart, and a 2-parameter family of $2\otimes n$ quantum system to verify the criteria given in Theorem 1.
\subsubsection{ A family of 4 $\otimes$ 4 NPT and bound entangled states }
Let us consider a family of $4\otimes4$ entangled states \cite{toth},\\
\begin{eqnarray}
	\rho_{p,q} = p \sum_{i=1}^4 |\psi_i\rangle \langle\psi_i| +  q \sum_{i=5}^6 |\psi_i\rangle \langle\psi_i|
\end{eqnarray} where $p$ and $q$ are non-negative real numbers satisfying $4p + 2q = 1$. $\{|\psi_i\rangle\}_{i=1}^6$ are defined as follows.

\begin{eqnarray*}
	&& |\psi_1\rangle = \frac{1}{\sqrt{2}} (|01\rangle + |23\rangle)\\
	&& |\psi_2\rangle = \frac{1}{\sqrt{2}} (|10\rangle + |32\rangle)\\
	&& |\psi_3\rangle = \frac{1}{\sqrt{2}} (|11\rangle + |22\rangle)\\
	&& |\psi_4\rangle = \frac{1}{\sqrt{2}} (|00\rangle - |33\rangle)\\
	&& |\psi_5\rangle = \frac{1}{2} (|03\rangle + |12\rangle) + \frac{|21\rangle }{\sqrt{2}}\\
	&& |\psi_6\rangle = \frac{1}{2} (-|03\rangle + |12\rangle)+ \frac{|30\rangle }{\sqrt{2}}
\end{eqnarray*}
Note a few important properties of the state $\rho_{p,q}$: \\
\textbf{P1.}
For this state, $det({\rho_{p,q}^R}^{\dagger} \rho_{p,q}^R) = 0$, which implies that the matrix obtained after applying the realignment operation on $\rho_{p,q}$ is not full rank.\\
\textbf{P2.} The state $\rho_{p,q}$ becomes invariant under partial transposition when $p = \frac{q}{\sqrt2}$ which implies that $\rho_{p_0,q_0}$ is a PPT state for $q_0 = \frac{\sqrt{2} - 1}{2}$ and $p_0 = \frac{1 - 2q}{4}$. \\
\textbf{P3.} It may be noted that $\|\rho_{p_0,q_0}^R\|_1 = 1.08579$, which is greater than 1. Thus, by the matrix realignment criterion, one can say that $\rho_{p_0,q_0}$ is a PPT entangled state. \\
Moreover, we find that $\rho_{p_0,q_0}^R$ has $8$ non zero singular values. Therefore, we have $k=8$. The degenerated characteristic equation is $ \lambda^8 + \sum_{i=1}^8 D_i \lambda^{8-i} = 0$ where $D_i's$ are defined in (\ref{dm}) and $D_8 = \prod_{i=1}^8 \sigma_i^2 (\rho_{p_0,q_0}^R)$. Thus, we find that in this example, the left hand side of the inequality (\ref{thm4}) is given by
$\mathcal{R}_1 \equiv 56 {D_8}^{1/8} + T_1 -1 = 0.02082 > 0$. So inequality (\ref{thm4}) is violated and Corollary-2 implies that the state $\rho_{p_0,q_0}$ is PPT entangled. We further observe that the criterion \cite{tzhang} given in (\ref{rzhang}) does not detect the BES described by the density operator $\rho_{p_0,q_0}$ belonging to the $\rho_{p,q}$ family.

On the other hand, when $(p, q) \neq (p_0, q_0)$, $\rho_{p,q}$ represents an NPT entangled state for which the detection range by employing our R-moment
criterion is given by  ($0.00659601 < q < 0.153105 $) and ($ 0.26477 < q \leq 1/2$), which is comparatively larger than the range ($0.425035 < q \leq 1/2 $)  detected by employing the $D_3^{(in)}$ criterion (\ref{d3}).

\subsubsection{3 $\otimes$ 3 NPT entangled state}
\begin{figure}[h!]
	\includegraphics[width=0.48\textwidth]{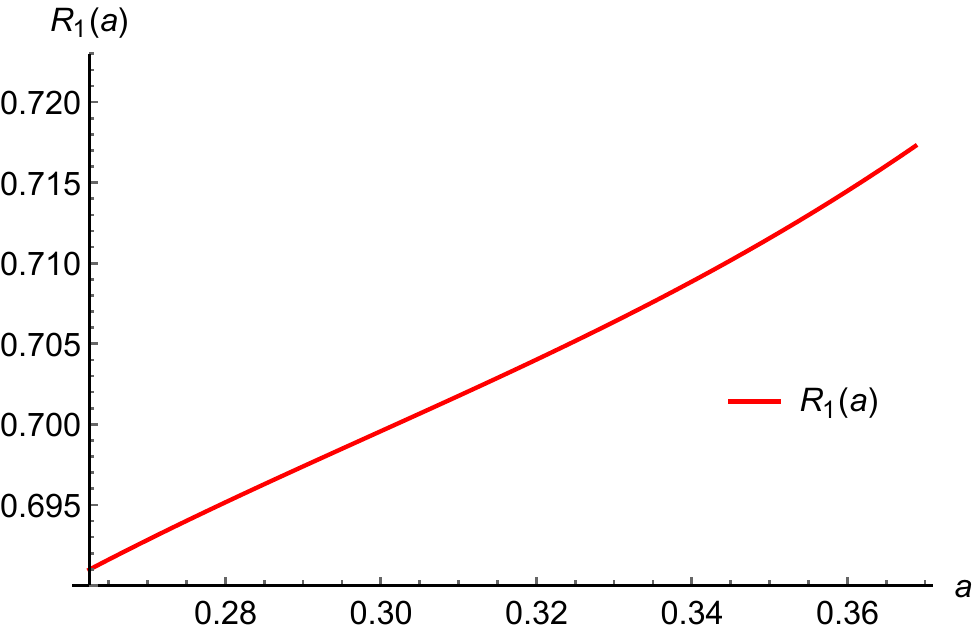}
	\caption{The red curve represents the $R$-moment criterion for the state $\rho_a$ given
		by Eq.(\ref{rhoa}). ${R}_1(a) > 0$ certifies the detection of entanglement in $\rho_a$. Here, $x$-axis represents the state parameter $a$. }
	\label{aimage1}
\end{figure}

\begin{figure}[h!]
	\includegraphics[width=0.48\textwidth]{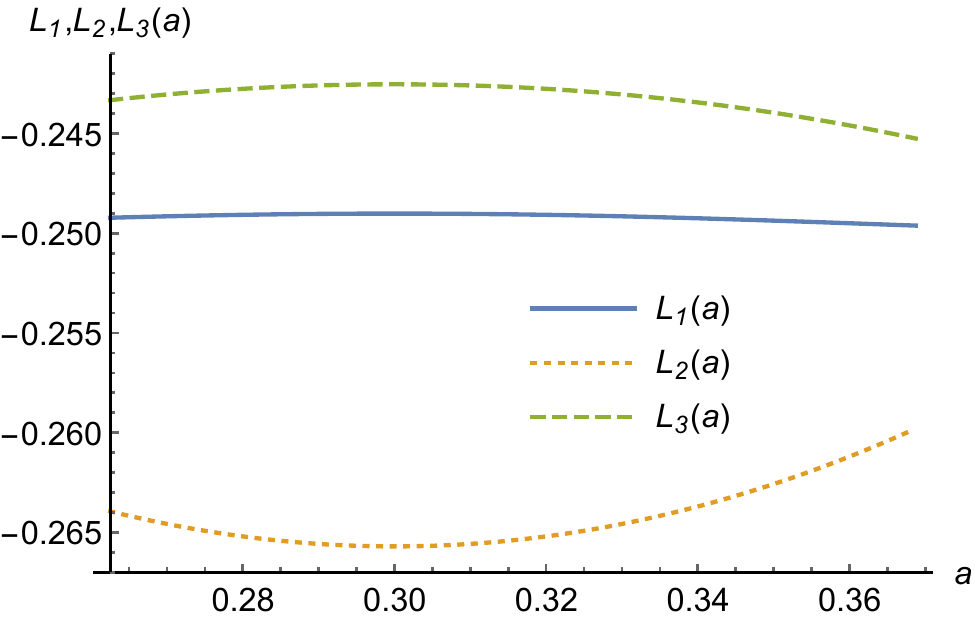}
	\caption{For the state $\rho_a$ (\ref{rhoa}), the blue (middle) curve represents the $p_3$-PPT criterion (\ref{p3ppt}), the dotted orange (lowermost) curve represents the $D_3^{(in)}$ criterion (\ref{d3}), and the dashed green (uppermost) curve represents the $p_3$-OPPT criterion (\ref{l3}). The graph is plotted with respect to the state parameter $a$. $L_1, L_2, L_3 < 0$ show that $p_3$-PPT, $D_3^{(in)}$, and $p_3$-OPPT criteria fail to detect this state in the whole range of $a$.}
	\label{aimage3}
\end{figure}

Next, consider the class of NPT entangled states in $3\otimes 3$ dimensional system, which is defined as
\cite{garg}
\begin{eqnarray}
	\rho_a =
	\begin{pmatrix}
		\frac{1-a}{2} & 0 & 0&0&0&0&0&0 & \frac{-11}{50}\\
		0 & 0&0&0&0&0&0 &0&0\\
		0 & 0&0&0&0&0&0 &0&0\\
		0 & 0&0&0&0&0&0 &0&0\\
		0 & 0&0&0&\frac{1}{2} - a&  \frac{-11}{50} &0 &0&0\\
		0 & 0&0&0&  \frac{-11}{50} &a&0 &0&0\\
		0 & 0&0&0&0&0&0 &0&0\\
		0 & 0&0&0&0&0&0 &0&0\\
		\frac{-11}{50} & 0&0&0&0&0&0 &0& \frac{a}{2}\\
	\end{pmatrix}
	\label{rhoa}
\end{eqnarray} 
where $\frac{1}{50} (25 - \sqrt{141}) \leq a \leq \frac{1}{100}(25 + \sqrt{141}) $. Since $\rho^R_a$ forms a matrix of rank $5$, so we have $k=5$. Thus, to detect whether $\rho_a$ is entangled or not, we need only 5 moments of $(\rho_a^R)^{\dagger} \rho_a^R$. Further we note that $T_1 = \frac{867}{1250} - \frac{3a}{2} + \frac{5 a^2}{2}$. 
By calculating all the five moments, we find that the inequality in (\ref{thm4}) is violated in the whole range of $a$. Thus, applying Corollary-1, we can say that $\rho_a$ is entangled for $\frac{1}{50} (25 - \sqrt{141}) \leq a \leq \frac{1}{100}(25 + \sqrt{141}) $.
Figure-\ref{aimage1} shows the violation of the inequality (\ref{thm4}), i.e., $ {R}_1 \equiv 20 {D_5}^{1/5} + T_1 - 1 > 0$ for all $0.262513 \leq a \leq 0.368743 $.
Figure-\ref{aimage3} shows the comparison of $R$-moment criterion with the $p_3$-PPT, $D_3^{(in)}$  and  $p_3$-OPPT criteria defined in (\ref{p3ppt}),  (\ref{d3}), and (\ref{l3}), respectively. Since  ${L}_1 < 0$, $ {L}_2 < 0$ and $ {L}_3 < 0$, we find that the state $\rho_a$ is not detected by any of the above partial moment based criteria for any value of the parameter $a$ in the given range. This is illustrated in Fig-\ref{aimage3}.

\subsubsection{Two-parameter class of states in
	$2 \otimes n$ quantum systems}
Consider the two-parameter class of state defined in $2 \otimes n$ quantum systems \cite{chi}:
\begin{eqnarray}
	\rho^{(n)}_{\alpha,\gamma} = \alpha \sum_{i=0}^{1}\sum_{j=2}^{n-1} |ij\rangle \langle ij| + \beta (|\phi^+\rangle \langle \phi^+| + |\phi^-\rangle \langle \phi^-| \nonumber\\ + |\psi^+\rangle \langle \psi^+|) + \gamma(|\psi^-\rangle \langle \psi^-|) \label{rhon}
\end{eqnarray}
where ${|ij\rangle; i=0,1; j=0, 1, . . ., n-1}$ forms an orthonormal basis for $2 \otimes n$ quantum systems,
\begin{eqnarray}
	|\phi^{\pm}\rangle = \frac{1}{\sqrt{2}} (|00\rangle \pm |11\rangle) ;\quad
	|\psi^{\pm}\rangle = \frac{1}{\sqrt{2}} (|01\rangle \pm |10\rangle)
\end{eqnarray}
By the trace condition, the parameter $\beta$ can be written in terms of $\alpha$ and $\gamma$ as
\begin{eqnarray}
	\beta = \frac{1 - 2(n-2)\alpha - \gamma}{3}
\end{eqnarray} 
where $0 \leq \alpha \leq \frac{1}{2(n-1)}$ and $0 \leq \gamma \leq 1$.

Now we compare the detection power of $R$-moment criterion with the moment based criterion given in Sec-II C. Let us consider the realignment moment based criterion given by Zhang et al \cite{tzhang} mentioned in (\ref{rzhang}), (\ref{hk}), and (\ref{bl}). Since in $2 \otimes n$ systems, $rank(\rho^{(n)}_{\alpha,\gamma}) \leq 4$, the criterion given in (\ref{rzhang}) is equivalent to the separability criterion based on Hankel matrices given in (\ref{hk}) and (\ref{bl}).

\begin{figure}[h!]
	\includegraphics[width=0.42\textwidth]{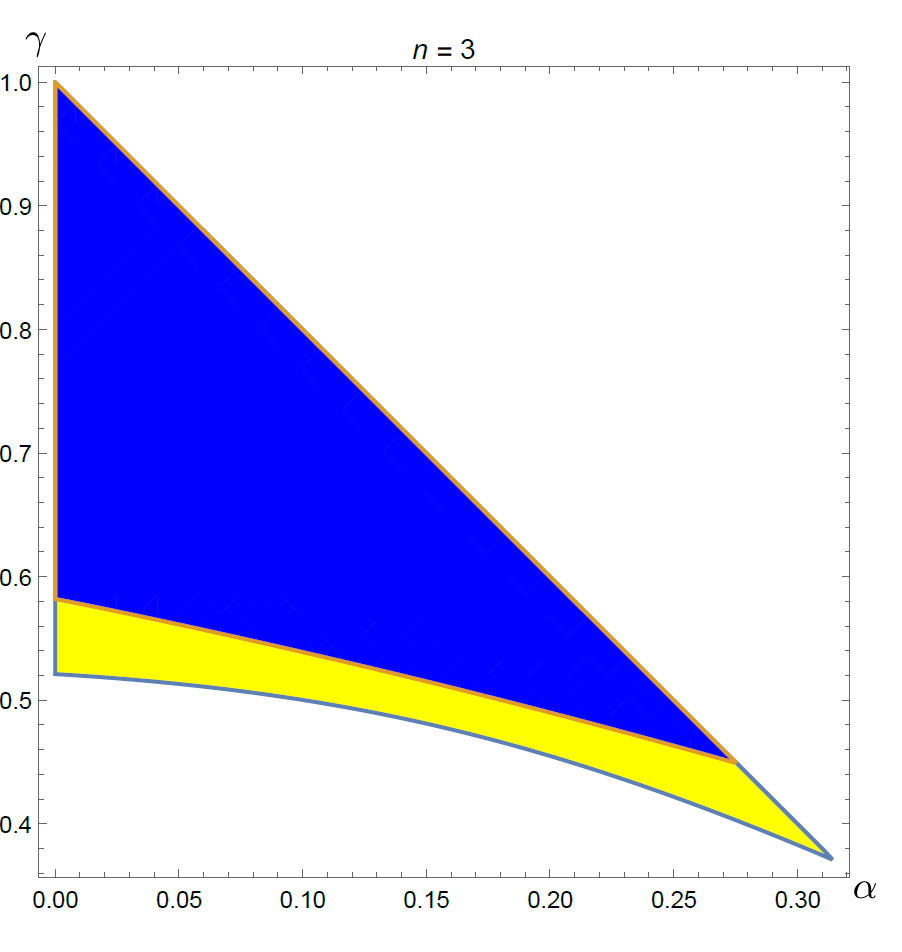}(i)\\
	\includegraphics[width=0.42\textwidth]{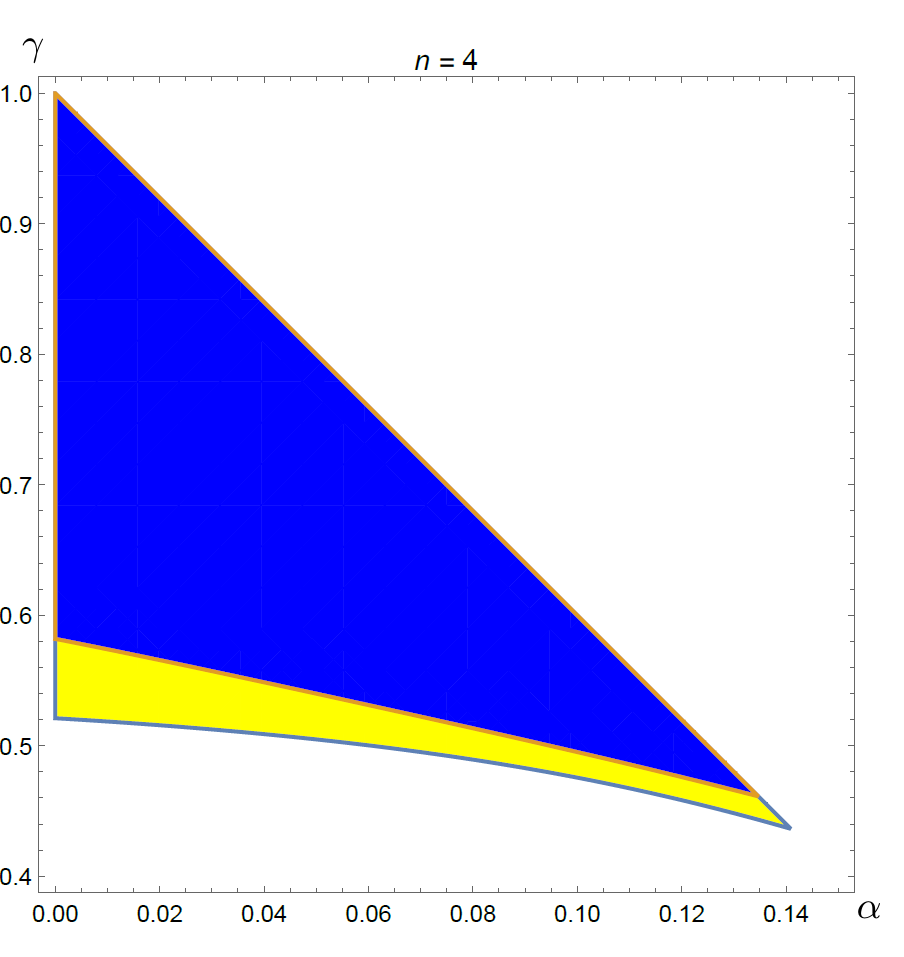}(ii)
	\caption{The region plot given above in (i) and (ii), respectively, shows the detection region of the state $\rho^{(n)}_{\alpha,\gamma}$, for $n = 3$ and $n=4$ by $R$-moment criteria and the Zhang's realignment moment based criteria given in (\ref{rzhang}).
		The blue (dark gray) region represents the entangled states detected by both the above mentioned criteria. The yellow (light gray) region shows the states detected by $R$-moment criteria but undetected by Zhang's criteria. Here, $x$-axis represents the state parameter $\alpha$ and $y$-axis represents the state parameter $\gamma$.}
	\label{alphagamma}
\end{figure}

In $2 \otimes 3$ systems, $\rho^{(3)}_{\alpha,\gamma}$ is entangled when $0 \leq \alpha \leq \frac{1}{4}$ and $\frac{1-2\alpha}{2} \leq \gamma \leq 1-2\alpha$. The inequality in (\ref{rzhang}) is violated, i.e., $L_4 (\alpha,\gamma)>0$ for $(\alpha, \gamma)$ lying in the blue (dark gray) shaded region in Fig-\ref{alphagamma}(i). Hence the entanglement is detected in this region by Zhang's criteria.
Now, applying the $R$-moment criterion on $\rho^{(3)}_{\alpha,\gamma}$, the inequality in (\ref{thm4}) is violated for the states lying in the blue (dark gray) as well as yellow (light gray) shaded regions. It can be thus seen that the R-moment 
criterion performs better for such systems. Similarly, it is also possible to show (see  Fig-\ref{alphagamma}(ii)) that for $n=4$ again a larger set of states is detected by the $R$-moment
criterion compared to the realignment criterion.

\section{Separability Criterion based on Realigned Moments ($R$-moments) in $2\otimes 2$ systems}
For $2\otimes 2$ dimensional systems the necessary and sufficient condition
for entanglement is provided by the Peres-Horodecki PPT criterion 
\cite{peres, horo}. It may be noted though that partial transposition
is positive but not completely positive, and hence, it is difficult 
to be realized directly in an experiment \cite{horoekert}. On the other hand, quantitative
measures of entanglement such as the entanglement of formation (EoF)
have been proposed \cite{woot1, woot2}, which have important applications
in using entanglement as a resource for implementing tasks such as
teleportation and dense coding \cite{woot3}. Bounds on the EoF have
further been proposed \cite{fei1, fei2}, which are in principle, measurable,
though precise measurement schemes for EoF are yet to be developed.
Note further, that certain other ingenious measurement schemes for
detecting two-qubit entanglement have been proposed, such as those based on
employing weak measurements \cite{entdet}. However, only pure states can be detected by local operations based on the weak measurement scheme.

Based on our analysis contained in the previous section, it can be shown that $D_4$ (\ref{thm5}) is non-zero for all 2-qubit entangled states (see Appendix A). Thus, we have to use full information of the entangled state to show the violation of the inequality (\ref{thm5}). 
This motivates us to develop a criterion that detects entanglement using less number of moments in $2 \otimes 2$ systems. 
In this section, we introduce a different separability criterion based on $R$-moments for $2 \otimes 2$ systems. The criterion is formulated in the form of an inequality that involves up to $3^{rd}$ order moments for detecting entanglement in a 2-qubit system. In general, any separability criterion for $2 \otimes 2$ system requires up to $4^{th}$ order moments, except for a few separability criteria \cite{elben,neven,guhne,lin} that require up to $3^{rd}$ order moments.  

\subsection{Prerequisites}
To develop the separability criteria for the two-qubit system, we need some prerequisites where we establish relations between moments of a matrix with its singular values or eigenvalues.\\
 It may not be an easy task to directly compute the eigenvalues of a matrix; thus, bounds for eigenvalues are of great importance.
Further, the bound of the eigenvalues expressed in terms of
moments may be useful for the experimentalist to estimate the
eigenvalues in the laboratory. Let us note the following results that give the lower bound and the upper bound of the largest eigenvalue of a complex matrix in terms of the first three moments of the matrix.\\
\textbf{Result 1:} Let $A$ be a complex $n \times n$ matrix with real eigenvalues $\lambda_i$ such that $\lambda_1 \geq \lambda_2. \; . \; . \geq \lambda_n$. If $T_k$ denotes the $k^{th}$ order moment of $A$ then, one 
has \cite{gupta} 
\begin{eqnarray}
	\lambda_1 \geq f(T_1, T_2, T_3) := \frac{T_1}{n} + \frac{b+ \sqrt{b^2 + 4 a^3}}{2a} \label{maxlb}
\end{eqnarray}
where
$$ a= \frac{T_2}{n} - \left(\frac{T_1}{n}\right)^2$$ and
$$ b= \frac{1}{n^3} (n^2 T_3 -3 n T_2 T_1 + 2 T_1^3) $$.\\
\textbf{Result 2:}  Let $A \in M_n(\mathbb{C})$ be a positive semi-definite matrix with eigenvalues $\lambda_1 \geq \lambda_2 . \; . \; . \geq \lambda_n$ and $T_k = Tr[A^k] $. Then the upper bound of the largest eigenvalue of $A$ is the largest root of the following cubic equation \cite{sharma}:
\begin{eqnarray}
	T_1 x^3 -2 T_2 x^2 + T_3 x + T_2^2 - T_1 T_3=0  \label{cubic}
\end{eqnarray}
i.e., 
\begin{eqnarray}
	\lambda_1 \leq g(T_1, T_2, T_3) \label{gtb}
\end{eqnarray}
where the upper bound $g(T_1, T_2, T_3)$ denotes the largest root of (\ref{cubic}).
The explicit form of $g(T_1,T_2,T_3)$ is given by
\begin{eqnarray}
	g(T_1,T_2,T_3):=
	\frac{1}{6T_1} (4 T_2 + \frac{2\times 2^{1/3} r}{(p+\sqrt{q})^{1/3}} + 2^{2/3} (p+\sqrt{q})^{1/3}) \nonumber\\ \label{gt}
\end{eqnarray}
where 
\begin{eqnarray*}
	&&	p = -27 T_1^2 T_2^2 +16 T_2^3 + 27 T_1^3 T_3 - 18 T_1 T_2 T_3\\
	&& r=  4T_2^2 -3 T_1 T_3\\
	&& q= p^2 - 4 r^3 
\end{eqnarray*}
Let $\rho$ be a density matrix representing a $2 \otimes 2$ dimensional state and $\rho^R$ denote the realigned matrix obtained after applying the realignment operation. Let $\sigma_{max}(\rho^R)$ be the largest singular value of $\rho^R$ and $\lambda_{max} ((\rho^{R})^{\dagger}\rho^{R}) $ be the largest eigenvalue of $(\rho^{R})^{\dagger}\rho^{R}$. Let  $\lambda^{lb}_{max} ((\rho^{R})^{\dagger}\rho^{R}) $ and $\lambda^{ub}_{max}((\rho^{R})^{\dagger}\rho^{R})$ denote respectively, the lower and the upper bound of $\lambda_{max} ((\rho^{R})^{\dagger}\rho^{R}) $. To develop our method, we use the lower and upper bound discussed in Result 1 and 2. \\ 
Using (\ref{maxlb}) and (\ref{gtb}) and since $\lambda_{i}((\rho^{R})^{\dagger}\rho^{R})= \sigma_{i}^2 (\rho^R)$, we have
\begin{eqnarray}
	\lambda_{max}^{lb}((\rho^{R})^{\dagger}\rho^{R}) \leq \sigma_{max}^2 (\rho^R) \leq \lambda_{max}^{ub} ((\rho^{R})^{\dagger}\rho^{R})
\end{eqnarray} 
where $\lambda_{max}^{lb} ((\rho^{R})^{\dagger}\rho^{R}) = f(T_1,T_2,T_3)$ and $\lambda_{max}^{ub} ((\rho^{R})^{\dagger}\rho^{R})= g(T_1,T_2,T_3)$ are functions of $T_1$, $T_2$ and $T_3$ for the matrix $(\rho^{R})^{\dagger}\rho^{R}$.\\
Now let us consider the following lemmas in which we derive few relations between moments of the matrix $(\rho^{R})^{\dagger}\rho^{R}$ and the singular values of $\rho^R$. We use these lemmas later to prove our main theorem.\\  
\textbf{Lemma 1:} If $\sigma_{i}$'s denote the singular values of the realigned matrix $\rho^R$ arranged in the descending order as $\sigma_1(\rho^R) \geq \sigma_2(\rho^R) \geq \sigma_3(\rho^R) \geq \sigma_4(\rho^R)$, then the following inequality holds:
\begin{eqnarray}
	\prod_{j=2}^{4} (\sigma_{1} (\rho^R) + \sigma_{j} (\rho^R)) 
	&\geq& \lambda_{max}^{lb} ||\rho^R||_1 + \sqrt{|D_3|}
	\label{lemma1}
\end{eqnarray}
where $\lambda_{max}^{lb}\equiv \lambda^{lb}_{max} ((\rho^{R})^{\dagger}\rho^{R})$ denotes the lower bound of the largest eigenvalue of $(\rho^{R})^{\dagger}\rho^{R}$. 
(See Appendix B for the proof.)\\
\textbf{Lemma 2:} If $\rho$ describes a $2 \otimes 2$ dimensional quantum state and $\rho^R$ denotes the realigned matrix of $\rho$, then we have
\begin{eqnarray}
	||\rho^R||_1^2 &\geq& 2 \sqrt{D_2} + T_1 \label{lemma2}
\end{eqnarray}
(See Appendix C for the proof.)\\
\textbf{Lemma 3:} If $\sigma_{i}$'s denote the singular values of the realigned matrix $\rho^R$ arranged in the descending order as $\sigma_1(\rho^R) \geq \sigma_2(\rho^R) \geq \sigma_3(\rho^R) \geq \sigma_4(\rho^R)$ and if $D_2$ and $T_1$ have their usual meaning, then we have the following equality:
\begin{eqnarray}
	\sum_{1< i < j} \sigma_{i}^2 (\rho^R) \sigma_{j}^2 (\rho^R)
	= D_2 - \sigma_{1}^2 (\rho^R)(T_1- \sigma_{1}^2 (\rho^R) ) \label{lemma3}
\end{eqnarray}
where $\sigma_{1}^2 (\rho^R)= \sigma_{max}^2 (\rho^R)$. (See Appendix D for the proof.)

\subsection{Criteria}
Now we are ready to discuss our separability criteria based on realigned moments for $2 \otimes 2$ systems.\\
\textbf{Theorem 2:} Let $\rho$  be a positive trace class linear operator acting on the Hilbert space $\mathcal{H}^2_A \otimes \mathcal{H}^2_B$. The realigned matrix $\rho^R$ has singular values arranged in the order as $\sigma_1(\rho^R) \geq \sigma_2(\rho^R) \geq \sigma_3(\rho^R) \geq \sigma_4(\rho^R)$. If $\rho_s$ denotes a separable state, then the following inequality holds:
\begin{equation}
	\mathcal{R}_2 \equiv \sqrt{3 X^{2/3} + 2 Y - 2T_1} - 1 \leq 0 \label{xy}
\end{equation}
where $X$ and $Y$ are the functions of the first three realigned moments, given by
$$X= \lambda_{max}^{lb} \sqrt{2\sqrt{D_2} + T_1} + \sqrt{|D_3|}$$ and $$ Y= T_1 - \lambda_{max}^{ub} + \sqrt{D_2 - \lambda_{max}^{ub} T_1 + (\lambda_{max}^{lb})^2}$$

\textit{Proof:} Let us start with the first realigned moment $T_1$. Using (\ref{e5}), it can be expressed as
\begin{eqnarray}
	T_1 &=& (\sum_{i=1}^{4} \sigma_{i} (\rho_s^R))^2 - 2\sum_{i < j} \sigma_{i} (\rho_s^R) \sigma_{j} (\rho_s^R)  \label{e51}
\end{eqnarray}
The second term of (\ref{e51}) can be expressed as
\begin{eqnarray}
	&&2\sum_{i < j} \sigma_{i} (\rho_s^R) \sigma_{j} (\rho_s^R) \nonumber\\ 
	&=&\sum_{i < j} (\sigma_{i} (\rho_s^R) + \sigma_{j} (\rho_s^R))^2 - \sum_{i < j} \sigma_{i}^2 (\rho_s^R)+ \sigma_{j}^2 (\rho_s^R) \\ 
	&=& \sum_{i < j} (\sigma_{i} (\rho_s^R) + \sigma_{j} (\rho_s^R))^2 -3T_1\\
	&=& \sum_{j=2}^4 (\sigma_{1} (\rho_s^R) + \sigma_{j} (\rho_s^R))^2 +  \sum_{1< i < j} (\sigma_{i} (\rho_s^R) + \sigma_{j} (\rho_s^R))^2 -3T_1 \nonumber \\ \label{sum}
\end{eqnarray}
Since, the arithmetic mean of a list of non-negative real numbers is greater than or equal to their geometric mean and $\sigma_i$'s for $i=1$ to $4$ are non-negative real numbers, we have 
\begin{eqnarray}
	\sum_{j=2}^4 (\sigma_{1} (\rho_s^R) + \sigma_{j} (\rho_s^R))^2 
	\geq 3(\prod_{j=2}^{4} (\sigma_{1} (\rho_s^R) + \sigma_{j} (\rho_s^R)))^{2/3} \label{amgm}
\end{eqnarray}
Using Lemma 1, the RHS of the inequality (\ref{amgm}) may be simplified to
\begin{eqnarray}
	3(\prod_{j=2}^{4} (\sigma_{1} (\rho_s^R) + \sigma_{j} (\rho_s^R)))^{2/3}
	\geq 3\left(\lambda_{max}^{lb} ||\rho_s^R||_1 + \sqrt{|D_3|}\right)^{2/3} \nonumber\\
	\label{eq28}
\end{eqnarray}
Using (\ref{eq28}) and Lemma 2 in (\ref{amgm}), we obtain
\begin{eqnarray}
	&&\sum_{j=2}^4 (\sigma_{1} (\rho_s^R) + \sigma_{j} (\rho_s^R))^2 \nonumber \\
	&&\geq 3 \left( \lambda_{max}^{lb} \sqrt{ 2 \sqrt{D_2} + T_1} + \sqrt{|D_3|} \right)^{2/3}\\ && = 3 X^{2/3} \label{X}
\end{eqnarray}
The second term of (\ref{sum}) can be expanded using Lemma 3 as
\begin{eqnarray}
	&&\sum_{1<i<j} (\sigma_{i} (\rho_s^R) + \sigma_{j} (\rho_s^R))^2 \nonumber \\ && = 2\sum_{i=2}^4 \sigma_{i}^2 (\rho_s^R)  + 2\sum_{1< i < j} \sigma_{i} (\rho_s^R) \sigma_{j} (\rho_s^R)\\
	&& \geq 2(T_1 - \sigma_{max}^2(\rho_s^R)) + 2\sqrt{ \sum_{1< i < j} \sigma_{i}^2 (\rho_s^R) \sigma_{j}^2 (\rho_s^R)}\\
	&& = 2(T_1 - \lambda_{max}^{ub}) + 2\sqrt{D_2 - \sigma_{max}^2 (\rho_s^R)(T_1 - \sigma_{max}^2 (\rho_s^R) )} \nonumber\\ \label{eq23}\\
	&& \geq  2(T_1 - \lambda_{max}^{ub}) + 2\sqrt{D_2 - \lambda_{max}^{ub} T_1 + (\lambda_{max}^{lb})^2 }\\
	&& = 2Y \label{Y}
\end{eqnarray}
Using (\ref{X}) and (\ref{Y}) in (\ref{sum}), we get
\begin{eqnarray}
	||\rho_s^R||_1 &\geq& \sqrt{3X^{2/3} + 2Y -2 T_1} \label{eq61}
\end{eqnarray}
Now we will use $||\rho_s^R||_1\leq 1$ in (\ref{eq61}) to obtain the desired result, i.e., if $\rho_s$ is separable, then 
\begin{eqnarray}
	\sqrt{3X^{2/3} + 2Y -2 T_1} \leq 1
\end{eqnarray}
Hence proved.\\ 
\textbf{Corollary-2:} 
If any two-qubit state $\rho$ violates the inequality (\ref{xy}), i.e., if $\mathcal{R}_2 > 0$, then the state is entangled.
\subsection{Examples}
\example
The two-qubit isotropic state is given by \cite{iso}
\begin{eqnarray}
	\rho_{f} = \frac{1-f}{3} I_2 \otimes I_2 + \frac{4f-1}{3} |\psi^+\rangle \langle \psi^+|,  0\leq f \leq 1
\end{eqnarray}
with 	$|\psi^+\rangle = \frac{1}{\sqrt{2}} (|00\rangle + |11\rangle)$ and $I_2$ denotes the $2\times2$ identity matrix.
One can easily verify that the isotropic state $\rho_f$ is entangled for $\frac{1}{2} < f \leq 1$ using the PPT and matrix realignment criteria, while the $D_3^{(in)}$ criterion given in (\ref{d3}) ensures that the state $\rho_f$ is entangled in the range $0.625< f \leq 1$. In order to apply the $R$-moment criterion on $\rho_f$, our task is to probe whether the inequality in (\ref{xy}) holds for $\rho_f$.
After simple calculations, we get 
\begin{eqnarray}
	&&T_1 = Tr[(\rho_f^{R})^{\dagger} \rho_f^R]= \frac{1}{3} (1 - 2 f + 4 f^2)\nonumber
\end{eqnarray}
Thus, for the isotropic state, the inequality (\ref{xy}) reads
\begin{equation}
	{R}_2 \equiv \sqrt{3X_f^{2/3} + 2Y_f -2 T_1} - 1 \leq 0 \label{iso}
\end{equation}
The above inequality is violated for $0.608594<f\leq 1$ and this implies that the state $\rho_f$ is entangled in this range which is better than that 
provided by the $D_3^{(in)}$ criterion.


\example
Consider the two-parameter family of $2 \otimes 2$ states represented by the density matrix \cite{rudolph}
\begin{eqnarray}
	\rho_{s,t} = 
	\begin{pmatrix}
		\frac{5}{8} & 0 & 0 & \frac{t}{2}\\
		0 & 0 & 0 & 0\\
		0 & 0 & \frac{1}{2}(s-\frac{1}{4}) & 0\\
		\frac{t}{2} & 0 & 0 & \frac{1-s}{2} \\
	\end{pmatrix}, t\neq 0, \frac{1}{4} < s \leq 1
\end{eqnarray} 
This state has non-positive partial transpose for all values of the state parameters $t$ and $s$ for which the state is defined. Therefore, by the PPT criterion, it is entangled for any non-zero value of the parameter $t$ and $\frac{1}{4} < s \leq 1$.

\begin{figure}[h!]
	\includegraphics[width=0.48\textwidth]{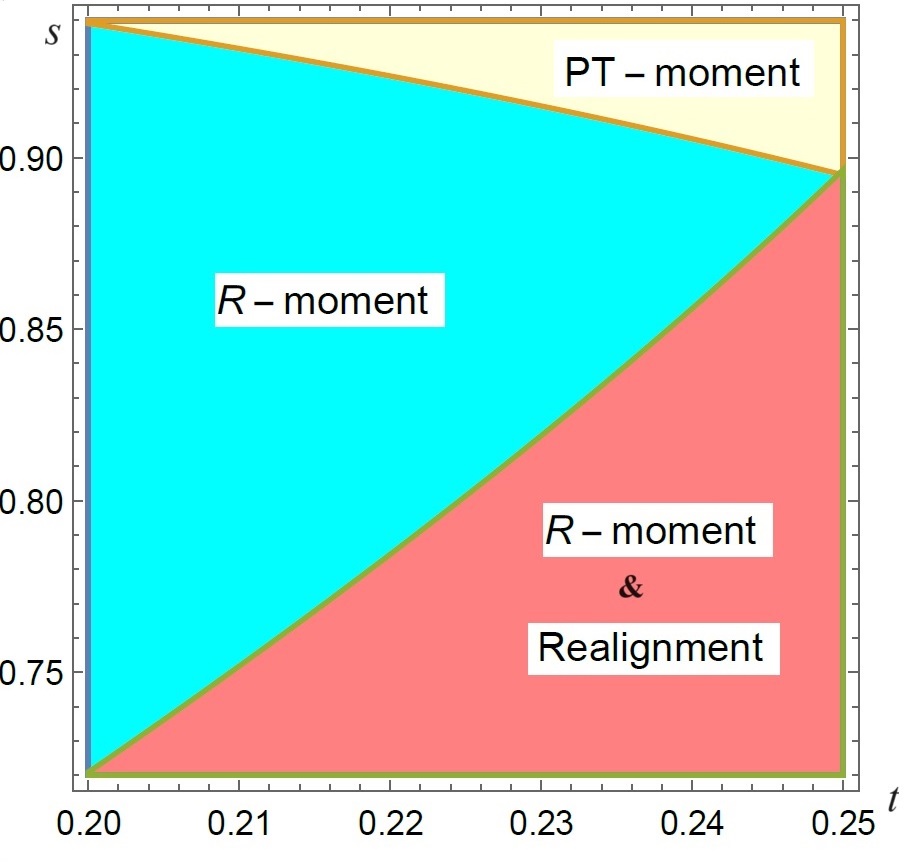}
	\caption{This figure shows the detection region of entangled states belonging to $\rho_{s,t}$ family of states. Here $x$-axis and $y$-axis denote the state parameters $t$ and $s$, respectively. The states lying in the blue (medium gray) region are detected by the $R$-moment criterion, but not by either the 
		criterion using partial transpose moments or by the realignment criterion. The states lying in the pink (dark gray) region are detected by $R$-moment as well as the realignment criterion. The yellow (light gray) region depicts the states detected by $p_3$-PPT as well as the $D_3^{(in)}$ criterion. } 
	\label{rud1}
\end{figure}

Let us now apply the $R$-moment criterion for the detection of the entangled state belonging to the family of the states described by the density operator $\rho_{s,t}$. The first moment of the Hermitian operator $(\rho_{s,t}^R)^{\dagger} \rho_{s,t}^R$ can be calculated as $T_1 = Tr[(\rho_{s,t}^R)^{\dagger} \rho_{s,t}^R]= \frac{1}{32}\left(21 -20s + 16( s^2 +|t|^2)\right)$. Similarly calculating $T_2$ and $T_3$ and putting these values in (\ref{xy}), we find that the state $\rho_{s,t}$ satisfies the inequality $\mathcal{R}_2 > 0$ for different ranges of the state parameter $s$ and $t$ as shown in Fig-\ref{rud1}. 
Further, to show the significance of $R$-moment criteria, we compare our criterion with partial transpose moments based criteria such as $p_3$-PPT and $D_{3}^{(in)}$ criterion given in (\ref{p3ppt}) and (\ref{d3}). It can be shown that the $D_{3}^{(in)}$ criterion performs better than the $p_3$-PPT criterion for detecting entangled states in the $\rho_{s,t}$ family.
\begin{eqnarray}
	{L}_2 > 0, \; \text{for}\;\; |t|> \sqrt{\frac{ 20s^2-25s +5 }{16s-26}}
\end{eqnarray}
It is important to note that the $R$-moment criterion detects those entangled states in $\rho_{s,t}$ family which are neither detected by partial transpose moments based criteria, such as $D_3^{(in)}$ and $p_3$-PPT criteria nor by the realignment criterion. This is illustrated in Fig-\ref{rud1} when $t \in [0.2,0.25]$. 
Similarly, we can get identical region $S=\{t  \;| -0.25\leq t \leq -0.2\}$  of entangled states detected by $R$-moment criterion but not by partial moment based criteria.\\
		
\section{Realization of moments of the realigned  matrix}
In this section, we formulate a procedure to show how the moments of the realigned matrix may be realized in an experiment.
Recently, it has been shown that the measurement of the moments of partially transposed density matrices is practically possible \cite{ha, cai, barti, saugato}. In this work, we consider the moments of the realigned matrix for its possible realization in an experiment. To achieve our aim, we adopt the idea presented in \cite{barti, saugato}, where it has been shown that the $k$th partial moment can be measured using SWAP operators \cite{kwek2002} on $k$ copies of the state. The technique involved is to express the matrix power as the expectation of the permutation operator. We adopt this approach applied to the realigned matrix in order to show how the measurement of realigned moments could be accomplished. 

For a $d\otimes d$ dimensional state $\rho$, the $k$ copies are given by $\otimes_{c=1}^k \rho_c$. Let $m_k$ denote the $k$th moment of the realigned matrix $\rho^R$, i.e., 
\begin{eqnarray}
	m_k = Tr[(\rho^R)^k] \label{mk}
\end{eqnarray}
The $k$th moment of $\rho^R$ can be expressed in terms of the expectation value of the permutation operator as 
\begin{eqnarray}
	m_k &=& Tr[(\otimes_{c=1}^k \rho^R_c) P^k] \label{mkc}
\end{eqnarray}
where $P$ is the normalized permutation operator defined as 
$P=\frac{1}{d} \sum_{i,j=0}^{d-1} |ij\rangle \langle ji |$ \cite{shruti5}. It is also known as the SWAP operator.

Since the separability criteria presented in Theorem-1 and Theorem-2 are based on the moments of the realigned matrix,  we need to estimate the moments of the matrix $(\rho^R)^\dagger \rho^R$. In particular, we show here the procedure of determining the first and second moments of $(\rho^R)^\dagger \rho^R$, which are denoted by $T_{1}$ and $T_{2}$, respectively.
Consider the following inequality that gives a relation between trace norm and Frobenius norm denoted by $||.||_{1}$ and $||.||_{F}$ respectively, of the realigned matrix $\rho^R$ \cite{zou}, given by
\begin{eqnarray}
	||\rho^R||_1 \leq \sqrt{k} ||\rho^R||_F
	\label{ineq1}
\end{eqnarray}
where $k$ is the rank of $\rho^R$.
The first moment of the realigned matrix is given by
\begin{eqnarray}
	m_1=	Tr[\rho^R] 
	\label{m100}
\end{eqnarray}
Applying the result $Tr[\rho^R] \leq ||\rho^R||_1$ and using the above inequality (\ref{ineq1}), we get 
\begin{eqnarray}
	m_1=	Tr[\rho^R] \leq \sqrt{k} ||\rho^R||_F = \sqrt{k \sum_{i=1}^k \sigma_{i}^2(\rho^R)} =\sqrt{kT_1}\nonumber\\
\end{eqnarray}
Hence, we have
\begin{eqnarray}
	T_1 \geq \frac{m_1^2}{k} \label{lbt1}
\end{eqnarray}
As the derived separability criterion depends on the singular values of the realigned matrix,  we further need to estimate the singular values of 
$\rho^R$. Here we use the result \cite{merikoski}
\begin{eqnarray}
	\sigma_{j}(\rho^R) \leq \frac{m_1}{d^2} + \sqrt{\frac{d^2-j}{j}\left(T_1 - \frac{m_1^2}{d^2}\right)} \label{sigub}
\end{eqnarray} 
where $1\leq j \leq k$.
The above inequality (\ref{sigub}) holds when 
\begin{eqnarray}
	T_1 \leq \frac{m_1^2}{j} \label{ubt1}
\end{eqnarray}
Now, combining (\ref{lbt1}) and (\ref{ubt1}), the first moment of $(\rho^R)^{\dagger}\rho^{R}$ is bounded from above and below by
\begin{eqnarray}
	\frac{m_1^2}{k} \leq T_1 \leq \frac{m_1^2}{j},~~1\leq j \leq k 
	\label{lbub}
\end{eqnarray}
It may be observed that the optimal value of $T_1$ may be obtained by substituting $j=k$ in (\ref{lbub}), which turns out to be
\begin{eqnarray}
	T_1^{opt} = \frac{m_1^2}{k}
\end{eqnarray}
Next, we obtain the estimate of $T_{2}$ for which we apply the following result.
For any two positive semi-definite matrices $A$ and $B$ in $M_n(\mathbb{C})$, the following matrix inequality holds \cite{yang},
\begin{eqnarray}
	(Tr[AB]^q) \leq (Tr[A])^q (Tr[B])^q
	\label{ineq2}
\end{eqnarray} 
where $q$ is a positive integer.
Now inseerting $A=B=(\rho^R)^{\dagger} (\rho^R)$ and $q=1$ in the above inequality, (\ref{ineq2}) reduces to
\begin{eqnarray}
	T_{2}=Tr[\left((\rho^R)^{\dagger} (\rho^R)\right)^2] \leq \left(Tr[(\rho^R)^{\dagger} (\rho^R)]\right)^2=(T_{1})^{2}
	\label{ineq3}
\end{eqnarray}
On the other hand, the lower bound turns out to be \cite{merikoski}
\begin{eqnarray}
	T_{2}\geq \frac{T_{1}^2}{d^2}
	\label{ineq4}
\end{eqnarray}
Combining (\ref{ineq3}) and (\ref{ineq4}), the bounds on $T_{2}$ can be expressed in terms of $m_{1}$, given by
\begin{eqnarray}
	\frac{m_{1}^4}{d^2k^2} \leq T_{2} \leq \frac{(m_{1})^{4}}{j^2},~~~1\leq j \leq k
	\label{ineq5}
\end{eqnarray}
The optimal range of $T_2$ may be obtained by putting $j=k$, which is given by
\begin{eqnarray}
	\frac{m_{1}^4}{d^2k^2} \leq T_{2}^{opt} \leq \frac{(m_{1})^{4}}{k^2}
	\label{ineqopt}
\end{eqnarray}
Therefore,  $T_{1}^{opt}$ and $T_{2}^{opt}$  may be experimentally determined
by measuring the first moment $m_1$ of $\rho^R$.

Since $\rho^R$ is not physically realizable, we need to first express it in terms of a physically realizable operator using an approximated map $\widetilde{R}$ called the structural physical approximation (SPA) \cite{horoekert}. The action of SPA map on realigned matrix $\rho^{R}$ may be defined as \cite{shruti3}
\begin{eqnarray}
	\widetilde{\rho^R}= \frac{p}{d^2} I_{d \otimes d} +\frac{(1-p)}{Tr[\rho^R]} \rho^R \;\; \text{where} \;\; 0 \leq p \leq 1 \label{65}
\end{eqnarray}
The approximated map $\widetilde{\rho^R}$ is positive as well as completely positive when  \cite{shruti3}
\begin{eqnarray}
	p \geq \frac{ d^2l}{m_1+ d^2l} 
	\label{p}
\end{eqnarray}
where $l=max[0,-\lambda_{min}^{lb}[\rho^R]]$ and $\lambda_{min}^{lb}[\rho^R]$ can be expressed as 
\begin{eqnarray}
	\lambda_{min}^{lb}[\rho^R]= \frac{m_1}{d^2} - \sqrt{(d^{2}-1)\left(\frac{m_2}{d^2}- \left(\frac{m_1}{d^2}\right)^2\right)}\nonumber\\ \label{lbr}
\end{eqnarray}
It may be observed from (\ref{65}) that the realigned matrix described by the density operator $\rho^R$ is proportional to its SPA operator $\widetilde{\rho^R}$, {\it i. e.},
\begin{eqnarray}
	\rho^R \propto \widetilde{\rho^R}- \frac{p}{d^2} I_{d \otimes d}
\end{eqnarray}
Hence, the $k$th moment of $\rho^R$ may be estimated as
\begin{eqnarray}
	m_k &\simeq& Tr	\left[\left(\otimes_{c=1}^k \left(\widetilde{\rho_c^R}- \frac{p}{d^2} I_{d \otimes d}\right) \right)P^k\right]\\
	&=& Tr	\left[\left(\otimes_{c=1}^k \widetilde{\rho_c^R} \right)P^m - \frac{p}{d^2} P^k \right]\\
	&=& Tr	\left[\left(\otimes_{c=1}^k \widetilde{\rho_c^R} \right)P^k\right] - \frac{p}{d^2} Tr[P^k] \label{mk} 
\end{eqnarray}
where $p \in [\frac{ d^2l}{m_1+ d^2l}, 1]$.
Since $\widetilde{\rho_{c}^R}$ is a Hermitian, positive semidefinite operator with unit trace,   $s_k := Tr	\left[\left(\otimes_{c=1}^k \widetilde{\rho_c^R} \right)P^k\right]$ can be measured using controlled swap operations \cite{horoekert, kwek2002}.

In particular, we show in Appendix E how the first moment $m_1$ may be 
determined. 
The first moment $m_1$ of $\rho^R$ may be estimated in terms of $s_1$ using the relations (\ref{case1}) and (\ref{case2}) given in Appendix E.  Since these relations are expressed in terms of $s_1= Tr[\widetilde{(\rho^R)}P]$, the first moment of realigned matrix $\rho^R$ can be estimated experimentally. 
It can be shown that the second moment of the realigned matrix can also be estimated similarly. 
Hence, the $k$th realigment moment may measured using (\ref{mk}).
Thus, this scheme can be generalized to higher dimensional systems as well. 
Hence, the measurement of the moments $m_k$ of the realigned matrix may be practically possible.

\section{Conclusion}
In this work, we have introduced a separability criterion for detecting the entanglement of arbitrary dimensional bipartite states based on partial information of the density matrix by employing realigned moments. Our proposed approach enables the detection of both PPT and NPT entangled states within the same framework using low ordered moments of the realigned matrix.
The formalism presented here is thus advantageous compared to the recently formulated entanglement detection schemes using partial transpose moments \cite{elben,neven,guhne} which fail to detect bound entangled states. 

We have demonstrated the significance of our separability criterion with the help of several examples of higher dimensional states.  We have compared the effectiveness of our criterion with other partial transpose moment based criteria. Our $R$-moment criterion for $m \otimes n$ systems is further significant since it can detect certain NPT entangled states that are undetected by the criteria based on partial transpose moments. Moreover, our separability criterion can detect
certain bound entangled states that are not detected by another recently
proposed criterion \cite{tzhang} using realigned moments.

Additionally, for two-qubit systems, we have shown that our approach can be
slightly modified to yield another separability criterion that can detect
entanglement in $2 \otimes 2$ systems without requiring complete information
about the quantum state. Interestingly, we have found that our realigned moment based criterion detects some two-qubit entangled states that are neither detected by partial moments based criteria \cite{elben,neven}, nor by the matrix realignment criteria \cite{chenwu, orud}. In \cite{liu}, authors have proposed a method based on permutation moments for the detection of multipartite entangled state. They have shown that their criterion reduces to partial transposition and realignment criterion in particular cases. In their criterion, odd order moments are inaccessible and their proposed criterion needs $2n$ copies of the entangled states to calculate $n$th order moment. On the other hand, odd and even moments are accessible in our criterion and it needs only $n$ copies of the state to calculate $n$th order moment. 

Finally, we have presented a scheme for the measurement of the moments of the
realignment matrix in order for our entanglement detection criterion to
be realized in practice. 
We conclude by noting that our proposed entanglement detection approach should be
experimentally implementable through the combination of techniques associated
with structural physical approximation for realignment \cite{horoekert, shruti3}, and SWAP operations for measuring density matrix moments \cite{barti, saugato}. \\

{\it Acknowledgements:}
 Shruti Aggarwal would like to acknowledge
the financial support by Council of Scientific and Industrial
Research (CSIR), Government of India (File no.
08/133(0043)/2019-EMR-1). ASM acknowledges support from the project No. 
DST/ICPS/QuEST/2018/98 of the Department of Science \& Technology, Government of India.



\section{Appendix}

\subsection{Non-existence of non-full rank realigned matrix of two-qubit entangled states}
\noindent 
In this section, we will show that the coefficient $D_{4}$ will not take value zero for any entangled two-qubit state. Let us consider an arbitrary two-qubit state that can be transformed by local filtering operation into either the Bell-diagonal state $\rho^{(BD)}=\sum_{i=1}^{4}p_{i}|\phi_{i}\rangle\langle \phi_{i}|$, where $|\phi_{i}\rangle's$ denote the Bell states, or the states described by the density operator $\rho^{(1)}$ which is of the form \cite{ishizaka} 
\begin{eqnarray}
	\rho^{(1)} = \frac{1}{2} 
	\begin{pmatrix}
		1+c & 0 & 0 & d\\
		0 & 0 & 0 & 0\\
		0 & 0 & b-c & 0\\
		d & 0 & 0 & 1-b \\
	\end{pmatrix}
\end{eqnarray}
where the state parameters $b$, $c$, $d$ satisfies any one of the following:\\
\begin{enumerate}
	\item[(C1)] $-1 \leq b < 1$, $~c=-1$, $~d=0$ 
	\item[(C2)] $b = 1$,$~-1< c \leq -1$, $~d=0$
	\item[(C3)] $-1 \leq b < 1$, $~-1< c \leq b$, $~d\leq |\sqrt{(1-b)(1+c)}|$ 
\end{enumerate}
\textbf{Case-I:} Let us consider the case when after the application of filtering operation on any two qubit state, the state is transformed as $\rho^{(1)}$. The state $\rho^{(1)}$ represents an entangled state if the following condition holds:
\begin{enumerate}
	\item[(E1)] $b,d \in R$, $~c \leq b$, $~d \neq 0$
\end{enumerate}
The realigned matrix of $\rho^{(1)}$ is denoted by $(\rho^{(1)})^{R}$ and it is given by
\begin{eqnarray}
	(\rho^{(1)})^{R} = \frac{1}{2} 
	\begin{pmatrix}
		1+c & 0 & 0 & 0\\
		0 & d & 0 & 0\\
		0 & 0 & d & 0\\
		b-c & 0 & 0 & 1-b \\
	\end{pmatrix}
\end{eqnarray}
The determinant of $(\rho^{(1)})^{R}$ is given by
\begin{equation}
	Det((\rho^{(1)})^{R})=\frac{(1-b)(1+c)d^{2}}{16}
	\label{det}
\end{equation} 
Using the conditions $(C1)$, $(C2)$, $(C3)$, and $(E1)$ in the determinant $Det((\rho^{(1)})^{R})$, it follows that the determinant must not be equal to zero. Thus, the hermitian matrix $((\rho^{(1)})^{R})^{\dagger}(\rho^{(1)})^{R}$ is a full rank matrix. Therefore $D_{4}\neq 0$ for any two-qubit entangled state $\rho^{(1)}$.\\
\textbf{Case-II:} If the state is transformed as a Bell diagonal state $\rho^{(BD)}$, then also it can be shown that the entanglement condition and $Det((\rho^{(BD)})^{R})=0$ does not hold simultaneously. This implies that in this case too  $D_{4}\neq 0$ for any two-qubit entangled state $\rho^{(BD)}$.
Thus, combining the above two cases, we can say that $D_{4}\neq 0$ for any two qubit entangled state.

\subsection{Proof of Lemma 1}
The LHS of inequality (\ref{lemma1}) can be expressed as
\begin{eqnarray}
	&&\prod_{j=2}^{4} (\sigma_{1} (\rho^R) + \sigma_{j} (\rho^R)) \nonumber \\&&= \sigma_{max}^2(\rho^R)||\rho^R||_1 + \sum_{i<j<k} \sigma_{i}(\rho^R)\sigma{j}(\rho^R)\sigma_k(\rho^R)\\
	&&\geq \lambda_{max}^{lb} ||\rho^R||_1 + \sqrt{\sum_{i<j<k} \sigma_{i}^2(\rho^R)\sigma_{j}^2(\rho^R)\sigma_k^2(\rho^R)} \label{e63} \\
	&&= \lambda_{max}^{lb} ||\rho^R||_1 + \sqrt{|D_3|}
	\label{eq50}
\end{eqnarray}
where the inequality (\ref{e63}) follows from the fact that  $(\sum_{i=1}^n x_i)^2 \geq \sum_{i=1}^n x_i^2$ holds for positive integers $x_i$, $i=1, 2, . . ., n$.\\

\subsection{Proof of Lemma 2}
The equation (\ref{e51}) can be re-written as
\begin{eqnarray}
	||\rho^R||_1^2 &=&	2\sum_{i < j} \sigma_{i} (\rho^R) \sigma_{j} (\rho^R) + T_1  \label{e65}
\end{eqnarray}

Applying the inequality $(\sum_{i,j=1}^n x_{ij})^2 \geq \sum_{i,j=1}^n x_{ij}^2$ in (\ref{e65}) for $x_{ij} = \sigma_{i} (\rho^R) \sigma_{j} (\rho^R)$,  we have
\begin{eqnarray}
	\sum_{i < j} \sigma_{i} (\rho^R) \sigma_{j} (\rho^R)  \geq \sqrt{\sum_{i < j} \sigma_{i}^2 (\rho^R) \sigma_{j}^2 (\rho^R)} \label{e66}
\end{eqnarray}

Using the inequality (\ref{e66}), the equation (\ref{e65}) reduces to
\begin{eqnarray}
	||\rho^R||_1^2 &\geq& 2 \sqrt{\sum_{i < j} \sigma_{i}^2 (\rho^R) \sigma_{j}^2 (\rho^R)} + T_1 \\ &=& 2 \sqrt{D_2} + T_1 \label{eq51}
\end{eqnarray}

\subsection{Proof of Lemma 3}
The LHS of (\ref{lemma3}) can be expressed as
\begin{eqnarray}
	\sum_{1< i < j} \sigma_{i}^2 (\rho^R) \sigma_{j}^2 (\rho^R) &=& \sigma_{2}^2 (\rho^R) \sigma_{3}^2 (\rho^R) + \sigma_{2}^2 (\rho^R) \sigma_{4}^2 (\rho^R)\nonumber\\&& + \sigma_{3}^2 (\rho^R) \sigma_{4}^2 (\rho^R)
 \\
	&=& D_2 - \sigma_{1}^2 (\rho^R)(\sum_{i=2}^4 \sigma_{i}^2 (\rho^R))\\
	&=& D_2 - \sigma_{1}^2 (\rho^R)(T_1- \sigma_{1}^2 (\rho^R) )
\end{eqnarray}

\subsection{Estimation of the first moment of the realigned matrix}

Eq.(\ref{mk}) may be re-expressed for $k=1$ as
\begin{eqnarray}
	m_1 &\simeq& Tr	[\widetilde{(\rho^R)}P] - \frac{p}{d^2} Tr[P]\\
	&=& Tr	[\widetilde{(\rho^R)}P] - \frac{p}{d^2}\\
	&\leq& Tr	[\widetilde{(\rho^R)}P] - \frac{l}{m_1+ d^2l} \label{m1}
\end{eqnarray}
We have used (\ref{p}) in the last step.
Therefore, the inequality (\ref{m1}) may be re-written as
\begin{eqnarray}
	m_1 + \frac{ l}{m_1+ d^2l} \leq  Tr	[\widetilde{(\rho^R)}P] := s_1 \label{s1}
\end{eqnarray}
Simplifying (\ref{s1}), we get 
\begin{eqnarray}
	m_1^2 + m_1(d^2l -s_1) + l(1-d^2s_1) \leq 0 \label{quad1}
\end{eqnarray}
Solving the above quadratic equation for $m_1$, we have
\begin{eqnarray}
	\frac{ -(d^2l - s_1) - \sqrt{(d^2l-s_1)^2 - 4l(1-d^2s_1)}}{2}	\leq m_1 \nonumber\\ \leq \frac{ -(d^2l - s_1) + \sqrt{(d^2l-s_1)^2 - 4l(1-d^2s_1)}}{2}
\end{eqnarray}
For $m_1$ to be real, we have 
\begin{eqnarray}
	(d^2l-s_1)^2 - 4l(1-d^2s_1) \geq 0\\
	\Rightarrow d^4l^2 + 2l(d^2s_1 -2) + s_1^2 \geq 0  \label{quad2}
\end{eqnarray}
Inequality (\ref{quad2}) holds when either $l \geq \frac{2 -d^2s_1+ 2\sqrt{1-d^2s_1}}{d^4}$ or $l \leq \frac{2 -d^2s_1 - 2\sqrt{1-d^2s_1}}{d^4}$\\
\textbf{Case 1:} When $ 2-d^2s_1+ 2\sqrt{1-d^2s_1} \leq d^4l \leq d^4 $
\begin{eqnarray}
	f_l(s_1)\leq	m_1 \leq f_u(s_1) \label{case1}
\end{eqnarray}
\textbf{Case 2:} When $0 \leq d^  4l \leq 2 -d^2s_1 - 2\sqrt{1-d^2s_1}$
\begin{eqnarray}
	g_l(s_1)\leq	m_1 \leq g_u(s_1) \label{case2}
\end{eqnarray}
where $f_l$, $f_u$, $g_l$, $g_u$ are functions of $d$ and $s_1$ given as follows:
\begin{eqnarray}
	f_l(s_1) &=&  \frac{1}{2}
	(-d^2 + s_1) \\&&  -\frac{1}{2d^2} \sqrt{d^8 + 2d^6s_1 + 4d^2s_1 + d^4s_1^2 -8(1+\sqrt{x})} \ \nonumber\\
	f_u(s_1) &=& \frac{-1}{d^2}(x + \sqrt{x}) + \nonumber\\&& \frac{1}{2d^2}\left(\sqrt{d^8 + 2d^6s_1 + 4d^2s_1 +d^4s_1^2 -8 (1+ \sqrt{x})}\right) \nonumber\\
	g_l(s_1) &=& \frac{1}{d^2}\left(  -x + \sqrt{x} - \sqrt{1+x-2\sqrt{x}} \right) \nonumber\\
	g_u(s_1) &=& \frac{s_1}{2} + \frac{1}{d^2} \sqrt{1+x-2\sqrt{x}}
\end{eqnarray}
where $x := 1 - d^2s_1$.

\end{document}